\documentclass[conference]{IEEEtran}
\IEEEoverridecommandlockouts

\usepackage{cite}
\usepackage{amsmath,amssymb,amsfonts}
\usepackage{algorithmic}
\usepackage{graphicx}
\usepackage{wrapfig}
\usepackage{lipsum}
\usepackage{textcomp}
\usepackage[dvipsnames]{xcolor}
\usepackage{url}
\usepackage{authblk}
\usepackage{pifont}

\usepackage{caption}
\DeclareCaptionFont{9pt}{\fontsize{9pt}{10pt}\selectfont}
\newcommand{\tool}[0]{\textit{CarbonClarity}}

\begin{document}

\title{CarbonClarity: Understanding and Addressing Uncertainty in Embodied Carbon for Sustainable Computing}



\author{Xuesi Chen, Leo Han, Anvita Bhagavathula, Udit Gupta}
\affil{Cornell Tech \\
       \texttt{\{xc562, lxh4, akb249, ugupta\}@cornell.edu}}

\maketitle

\begin{abstract}
Embodied carbon footprint modeling has become an area of growing interest due to its significant contribution to carbon emissions in computing. However, the deterministic nature of the existing models fail to account for the spatial and temporal variability in the semiconductor supply chain. The absence of uncertainty modeling limits system designers’ ability to make informed, carbon-aware decisions.
We introduce \tool, a probabilistic framework designed to model embodied carbon footprints through distributions that reflect uncertainties in energy-per-area, gas-per-area, yield, and carbon intensity across different technology nodes. Our framework enables a deeper understanding of how design choices, such as chiplet architectures and new vs. old technology node selection, impact emissions and their associated uncertainties. For example, we show that the gap between the mean and 95$^{th}$ percentile of embodied carbon per cm² can reach up to 1.6$\times$ for the 7nm technology node. Additionally, we demonstrate through case studies that: (i) \tool\text{} is a valuable resource for device provisioning, help maintaining performance under a tight carbon budget; and (ii) chiplet technology and mature nodes not only reduce embodied carbon but also significantly lower its associated uncertainty, achieving an 18\% reduction in the 95$^{th}$ percentile compared to monolithic designs for the mobile application. 

\end{abstract}


\section{Introduction}

\begin{figure}[h]
    \centering
    \includegraphics[width = \linewidth]{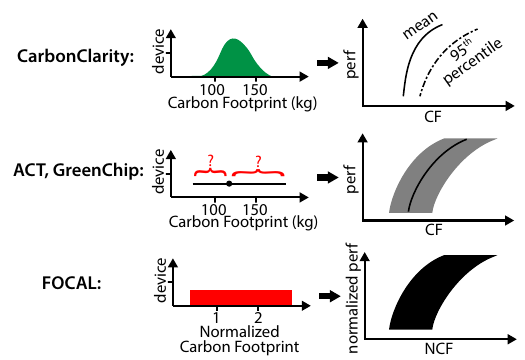}
    \vspace*{-6mm}   
    \caption{\tool~incorporates sources of uncertainty into embodied carbon modeling to generate probabilistic distributions of quantitative embodied carbon footprints. In contrast, prior tools such as ACT\cite{act} and GreenChip\cite{greenchip} provide deterministic point estimates without offering quantitative bounds on uncertainty. FOCAL~\cite{focal}, on the other hand, considers uncertainty through a comparative analytical model that constructs a uniform uncertainty range but relies on designers having prior knowledge of whether the design is operational or embodied carbon dominated. By modeling uncertainty explicitly, \tool~provides quantitative statistical insights that can more accurately guide computing system and hardware design compared to prior works~\cite{act,greenchip,focal}.}
    \label{fig:motivating}
\end{figure}

The modern Information and Communication Technology (ICT) industry accounts for 1.8-2.8\% of worldwide greenhouse gas (GHG) emissions~\cite{freitag2021real}, on par with the aviation industry. The environmental impact of computing technology is only expected to grow and even double by 2030 due to its increasing demand~\cite{sustainableAIpaper, chasingcarbon}.
Mitigating the environmental footprint of computing requires specialized solutions and designs to enable sustainable computer systems~\cite{chasingcarbon, act, ecochip}.
However, mitigating the environmental impact of computing poses fundamentally new challenges.
First, while hardware and software engineers have enabled tremendous optimizations in energy efficiency, performance and power over the past two decades, operational use is no longer the dominant source of emissions.
For example, recent efforts characterizing the carbon emissions of computing platforms throughout the lifetime of the hardware show \textit{embodied carbon}, owing to hardware manufacturing and fabrication, account for 75\% of the emissions in mobile devices over their lifetime~\cite{greenchip, chasingcarbon, act, apple2024report}; \textit{operational emissions}, owing to energy consumed during the device's lifetime amounts to roughly 20\% of lifetime emissions.
Second, optimizing the carbon footprint of computing hardware requires understanding and quantifying the sources of emission by navigating the uncertainty inherent in supply chains, manufacturing processes, and use cases.
Unfortunately, while various tools and methodologies exist to quantify system performance, power, and energy, the quantification of embodied emissions, especially with consideration of their underlying uncertainty, remains a nascent effort~\cite{act, greenchip,focal, ecochip}.

In particular, existing carbon models for computer systems are affected by significant uncertainty and noise arising from spatio-temporal variability in data collection, large error margins, or simply under-reporting. For example, researchers showed a range of more than 4$\times$ between storage emission rates (i.e., $GB$ per $kg\ CO_2$) in publicly available industry product environmental reports~\cite{dirtySecretSSD}.
Similarly, iMec device characterization between 2020 and 2023 shows up to a 1.45$\times$ deviation~\cite{imec2020, imec2023} for carbon emission of technology nodes.
These data inconsistencies are not necessarily erroneous; many sources of uncertainty exist across the supply chain including fab location, fabrication energy and process efficiency, and manufacturing yield~\cite{supplyChain}.
Companies use carbon estimates to guide sustainability programs and ensure devices continue to reduce emissions across generations to meet science-based climate targets~\cite{apple2024report, googleEnvironmentalReport}; the significant uncertainty in carbon estimates forces designers to make conservative, worst-case estimates~\cite{apple2024report, googleEnvironmentalReport}.
In fact, Google's corporate sustainability report states ``understanding the sources, types, and magnitude of uncertainty is crucial to deploy conservative estimates, inform improved data inputs, and properly interpret results''.


In this paper, we propose \tool, a probabilistic framework design to enable hardware designers to quantify, understand, and mitigate uncertainty in estimating the embodied carbon.
\tool\text{}~quantifies uncertainty in embodied carbon estimates by holistically modeling variability in key supply chain and fabrication parameters, including spatiotemporal variability in carbon intensity of energy in fabs, energy efficiency of fabs, abatement strategies, and yield.
\tool\text{}~addresses a crucial need to make robust, sustainable hardware. 
To mitigate uncertainty in carbon accounting, \tool~uses ~95$^{th}$ percentile to quantify \textit{conservative} carbon estimates\cite{SMC,quantileRegression,statJavaPerfEval}.
To demonstrate the efficacy of the uncertainty modeling approach, we integrate \tool\text{}~with the Architectural Carbon Modeling Tool, ACT~\cite{act}, an open-source hardware carbon model that has been used in Meta's internal carbon accounting toolchains~\cite{metaSustainabilityReport} and in various academic and industry research proposals~\cite{greenSKU, 3dcarbon, 3D-IC, scarif, focal}.

Using \tool\text{}~we conduct a series of evaluations and case studies.
First, we demonstrate that compared to state-of-the-art architectural carbon models such as ACT and FOCAL~\cite{act, focal}, \tool\text{}~provides hardware designers with a finer-grained understanding of the uncertainty in the carbon footprint of devices; \tool\text{}~provides hardware designers with an easier-to-use interface than prior work by not requiring a priori knowledge of operational to embodied carbon ratios of devices and instead allowing the designer to define arbitrary distributions for all inputs.
Second, we show that designers can use \tool\text{}~to balance performance and uncertainty in carbon by using chiplet design strategies.
Finally, we illustrate how \tool\text{}~can be used as a diagnostic tool to understand the key sources of uncertainty in hardware manufacturing.
Altogether, \tool\text{}~will enable hardware designers to make informed and risk-aware decisions that lead to robust sustainable designs, ultimately helping achieve carbon neutrality.





The main contributions of this work are: 

\begin{enumerate}
    \item We propose a novel framework that generates probabilistic distributions for embodied carbon estimates, leveraging publicly available data from key fabrication steps. The framework empowers hardware designers to make informed and robust sustainable design choices.
    We show that across 28nm to 7nm, the gap between average and 95$^{th}$ percentile embodied carbon, in terms of CO$_2$ per cm$^2$ of die area, is up to 1.6x (Section III). 
    
    \item We demonstrate that using the ~95$^{th}$ percentile carbon estimate from the proposed framework results in a 34\% performance improvement compared to the current conservative worst-case carbon estimation. Additionally, it ensures that the risk of exceeding the target budget remains below 5\% for AI accelerator resource provisioning (Section IV).

    \item We illustrate how chiplet-based chip design not only reduce average but also uncertainty in embodied carbon. For instance, for server-level CPU, compared to monolithic chip design, chiplets achieve up to 39\% and 44\% carbon reduction for average and 95$^{th}$ percentile, respectively (Section V).
    
    \item To mitigate the uncertainty in embodied carbon emissions, we diagnose the degree of uncertainty of various contributing fabrication parameters. 
    We show that using renewable resources can help reduce the overall uncertainty on embodied emission due to high variability on energy-per-area(EPA) values (Section V). 
\end{enumerate}

\tool~is available in the below GitHub repository: 
\url{https://github.com/S4AI-CornellTech/CarbonClarity}

\section{Background and Related Works}
Carbon emissions of computing can be categorized into two types: operational and embodied.
Operational carbon refers to the carbon emission that occurred during hardware use---the combination of energy consumption and carbon intensity of energy source powering the device. 
Embodied carbon refers to the carbon emission that occurred during the chip manufacturing process~\cite{chasingcarbon, act}
With the growing use of renewable energy~\cite{chasingcarbon}, the primary source of ICT’s carbon emissions is shifting from operational activities to manufacturing for both cloud and edge computing~\cite{act, greenSKU, apple2024report}.

\textbf{Carbon modeling:} Several existing carbon modeling tools have quantified the carbon footprint for individual IC’s. ACT\cite{act} introduces an architectural carbon footprint modeling framework to support sustainability-driven early design space exploration. ACT shows that the embodied carbon footprint of hardware includes various aspects of chip production in the fabrication facilities (fabs), from the procurement of raw materials to the energy consumption and greenhouse gas (GHG) emissions involved. For each of the application processors produced, the total embodied carbon emissions ($\text{E}_{SoC}$) are a function of the die area (Area) and the carbon emitted per unit area manufactured (CPA), shown in Fig. \ref{fig:framework}.



ECO-CHIP\cite{ecochip} examines the impact of heterogeneous integration on the embodied carbon emissions of chiplet-based systems. FOCAL~\cite{focal}, SCARIF~\cite{scarif}, and Greenchip~\cite{greenchip} extend carbon modeling by integrating it with architecture simulations, enabling analysis of the sustainability impact across various architectural scenarios, such us: mobile vs datacenter, as well as compute-to-memory trade-offs.

However, existing state-of-the-art embodied carbon modeling tools lack consideration of uncertainty across various manufacturing stages and supply chains. Chip manufacturing is a complex and iterative process influenced by numerous factors, including variations in fabrication locations, energy sources, process efficiencies, material sourcing, and manufacturing yields. These parameters can fluctuate significantly across the supply chain, introducing substantial uncertainty that current models fail to adequately capture, as highlighted by Supply Chain Aware Computer Architecture~\cite{supplyChain}.

\textit{We propose \tool\text{}~to address these challenges and helps quantify and reduce uncertainty in embodied carbon estimates. By modeling variations in factors like carbon intensity, fab efficiency, abatement strategies, and yield, \tool\text{}~enables hardware designers to make more sustainable decisions.}

\textbf{Statistical Approaches toward Uncertainty:} A few prior works have use statistical tools to quantify the uncertainty that arise in architecture and carbon modeling~\cite{SMC, U-DUCT, architectureRisk}. 
For example, SMC~\cite{SMC} has used statistical model checking to computer processors to account for the experimental variability and probabilistic evaluation of computer architecture properties of interest. 

\textit{\tool\text{}~differs from the above work in doing a detailed modeling of spatio-temporal uncertainty in the embodied carbon process.
}
\section{CarbonClarity: Modeling Uncertainty in Embodied Carbon}

\begin{figure}
    \centering
    \includegraphics[width = \linewidth]{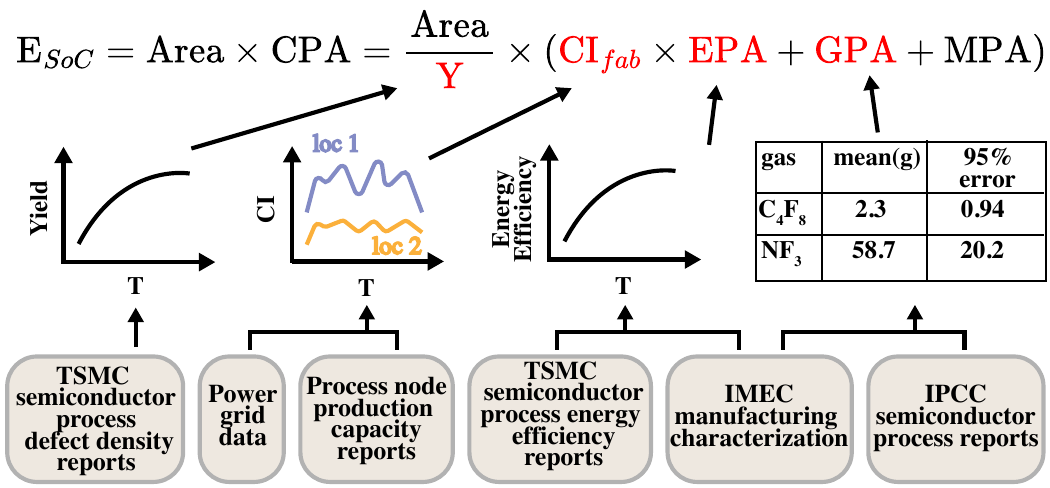}
    \vspace*{-5mm}
    \caption{An overview of the probabilistic modeling framework. The embodied carbon of the SoC is influenced by uncertainties in yield, CI$_{fab}$, EPA, and GPA, captured in various spatio-temporal formats. These uncertainty data are extracted from publicly available reports to construct the probabilistic modeling framework.} 
    \label{fig:framework}
    \vspace{-1mm}
\end{figure}

In this section, we describe \tool, an embodied carbon model that incorporates the underlying uncertainty to guide robust system design.
Fig.~\ref{fig:framework} illustrates the high-level process by which we augment existing architectural carbon modeling tools to consider uncertainty.
\tool~incorporates uncertainty in four key parameters: 
\begin{itemize}
    \item fabrication yield (Y): percentage of defect-free chips produced;
    \item carbon intensity of fab power grids (CI$_{fab}$): the carbon emissions per unit of electricity used by semiconductor fabs;
    \item energy per unit die area (EPA): energy consumed per unit
area during manufacturing;
    \item chemicals/gasses per unit die area (GPA): amount of GHG emissions (e.g. C$_2$F$_6$, NF$_3$, CF$_4$, SF$_6$, N$_2$O) released per unit area during manufacturing.
\end{itemize}

While previous architectural carbon models (e.g., ACT~\cite{act}, ECO-CHIP~\cite{ecochip}, GreenChip~\cite{greenchip}) generate point estimates per technology node for the ``average'' carbon-per-unit-area (CPA), \tool~generates distributions.
We generate distributions of carbon estimates based on spatio-temporal variance of each of the parameters using publicly available device characterization, industry sustainability reports, or reports from the International Panel on Climate Change\cite{ipcc2006,ipcc2019,TSMCepa,defectDensity,imec2020,imec2023,electricity_maps_south_korea_2024,globalCapacityBreakdown}.

Because available data for EPA, GPA, yield, and CI$_{fab}$ are typically discrete, sparse, and do not follow an obvious underlying distribution, we apply kernel density estimation (KDE) with Gaussian kernels~\cite{kde} to create probability density functions (PDFs) for each parameter. KDE is particularly suited for this task because it offers a combination of non-parametric flexibility, smoothing properties, and the ability to apply weights to individual data points. This contrasts with approaches used in prior work, where simple averaging or case-based analysis was employed~\cite{act,focal}. KDE enables us to transform finite, discrete data points into continuous probabilistic distributions without assuming a specific data model, providing a more robust treatment of uncertainty.

Moreover, the ability to apply weights to kernels is critical for integrating global manufacturing distribution information with regional carbon intensity data. For instance, we can assign higher weight to regions with larger semiconductor production shares, ensuring that regional variations in CI$_{fab}$ are appropriately reflected in the final carbon uncertainty model, as illustrated in Fig. \ref{fig:epa}.

Specifically, Fig.~\ref{fig:epa} shows the process of generating the EPA distribution for 16nm node in two stages: the left plot illustrates how Gaussian kernels are applied to individual data points, while the right plot demonstrates how these kernels combine to form a smoothed probability distribution. After generating the individual parameter-level distributions for EPA, GPA, yield, and CI$_{fab}$, we compute the full uncertainty distribution for carbon-per-unit-area (CPA) by combining these distributions through outer sum or product operations. The resulting distribution captures variability across the entire semiconductor manufacturing process and forms the probabilistic foundation of \tool's embodied carbon modeling framework.

In the following sections, we first describe the construction of individual parameter distributions, then present the end-to-end uncertainty modeling enabled by \tool.

\subsection{Parameter-Level Uncertainty Characterization}
We begin by studying the parameter-level uncertainty characteristics for EPA, GPA, yield, and CI$_{fab}$.

\subsubsection{GPA Distribution}

\begin{figure}
    \centering
    \includegraphics[width = \linewidth]{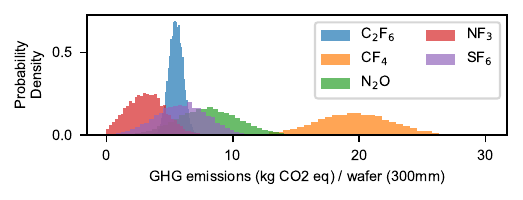}
    \vspace*{-8mm}
    \caption{Various greenhouse gases distribution for 14nm technology node}
    \label{fig:gpa}
    \vspace{-4mm}
\end{figure}

Gas per area (GPA) refers to the quantity of greenhouse gases released per unit area (cm$^2$), commonly occurring from Tier 2 activities, such as dry etching and chemical vapor deposition (CVD). While many gases, such as SF$_6$ and NF$_3$ do not directly contain carbon, they have high global warming potentials, resulting in large CO$_2$-equivalent emissions. 
The current architecture carbon models (e.g. ACT~\cite{act}) uses mean value with 95\% and 99\% abatement from imec reports~\cite{imec_gpa, imec2020, imec2023}. However, IPCC reports show that Tier 2 activity exhibit significant uncertainty, with SF$_6$ showing 300\% relative error rate~\cite{ipcc2006}. Therefore, we account the GPA uncertainty by applying the 95\% Tier 2 abatement errors provided by IPCC reports ~\cite{ipcc2006, ipcc2019} assuming normal distribution. Fig. \ref{fig:gpa} shows the carbon equivalent emission distribution for various gases occurred during the manufacturing process. 

\subsubsection{EPA}
\begin{figure}
    \centering
    \includegraphics[width = \linewidth]{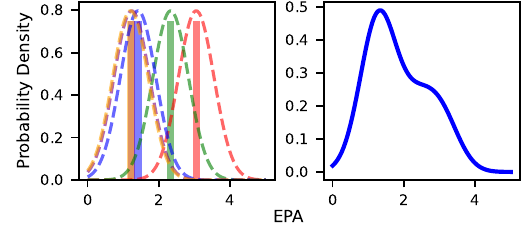}
    \vspace*{-6mm}
    \caption{The process of constructing a probability density using KDE for 16nm technology node based on its EPA value from 2015 to 2019. The left plot shows individual Gaussian kernels generated from each temporal EPA data point. The right plot displays the resulting probability density curve, obtained by averaging the sum of all kernels.}
    \label{fig:epa}
    \vspace{-1mm}
\end{figure}

EPA represents the energy consumed per unit area during manufacturing, derived from fabrication energy data~\cite{TSMCepa,imec2023,imec2020}. 
The uncertainty in EPA arises from variations in process energy efficiencies over time. For example, the energy efficiency of 28nm technology improved by 2.6$\times$, 3 years into mass production~\cite{TSMCepa}. 
To construct the EPA uncertainty distributions, we apply TSMC’s annual process energy efficiency improvements~\cite{TSMCepa} to the raw EPA values published on the device-level carbon characterization works~\cite{imec2020, imec2023}. This results in an EPA value for each year since mass production for a given technology node. 
Each yearly EPA value serves as a kernel in the KDE, and the Gaussian distributions for each kernel are averaged and summed to form the overall EPA uncertainty distribution for each technology node, shown in
Fig. \ref{fig:epa}. 

\subsubsection{Yield Distribution}
\begin{figure}
    \centering
    \includegraphics[width=\linewidth]{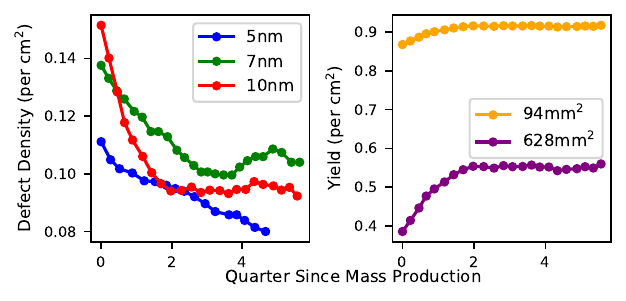}
    \vspace*{-7mm}
    \caption{The left figure shows the decrease of defect density per cm$^2$ for 5nm, 7nm and 10nm nodes  since mass production ~\cite{defectDensity}. The right figure shows two 10nm chips with different areas can exhibit varying yields over time. }
    \label{fig:yield}
    \vspace{-1mm}
\end{figure}

Yield refers to the percentage of defect-free chips produced on a semiconductor wafer relative to the total number of fabricated chips. As the fabrication technology for a given node matures, defect densities decrease, leading to an increase in yield variations in defect densities over time since mass production. For example, the per cm$^2$ defect density for 10nm chip dropped by 6\% in the first half year since mass production ~\cite{defectDensity}.
We utilized TSMC’s defect density ($D_0$) across three different process nodes~\cite{defectDensity}. The Poisson model, $\text{Yield} = e^{-D_0A}$, is used to calculate the chip's yield distribution, based on its area (A) and the defect density  of the technology node used~\cite{poisson}. Fig. \ref{fig:yield} shows the yield uncertainty distribution of dies with various area sizes.

\subsubsection{Carbon Intensities Distribution}

\begin{figure}
    \centering
    \includegraphics[width=\linewidth]{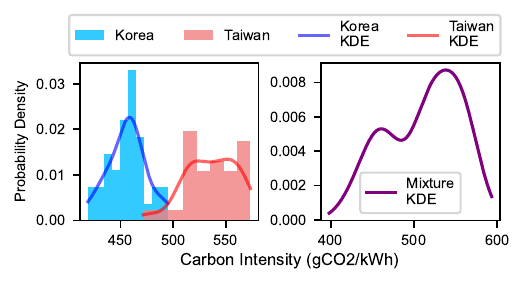}
    \vspace*{-8mm}
    \caption{The left figure displays carbon intensity uncertainty distribution for both Korea and Taiwan between 2021-2023 ~\cite{electricity_maps_south_korea_2024,electricity_maps_taiwan_2024}. The right figure showcases a weighted combination of two, forming a carbon intensity uncertainty distribution for 10nm technology node chip with unknown fab location.}
    \label{fig:ci}
    \vspace{-5mm}
\end{figure}

Carbon intensities of fab, $\text{CI}_{fab}$, represents the amount of carbon dioxide equivalent emissions generated per unit of electricity consumed by semiconductor fabrication facilities.
The geographical location of the fab, as well as the daily and seasonal fluctuations of renewable resources in the electric grid mix, contribute to uncertainty in $\text{CI}_{fab}$. 
If the fab location cannot be identified, we construct the uncertainty distribution $\text{CI}_{fab}$ based on three years of historical carbon intensity data from Electricity Maps~\cite{electricity_maps_south_korea_2024,electricity_maps_taiwan_2024}
For chips with unknown fabrication location, we determine the breakdown of global production capacity for the given technology node and use the percent of capacity manufactured in a country as the weight for the KDEs. We then aggregate the weighted kernels of each country's historical carbon intensity distribution to form the $\text{CI}_{fab}$ distribution for that node. 
For example, for a 10nm process node, the global production capacity in 2022 was split between South Korea (31\%) and Taiwan (69\%) ~\cite{globalCapacityBreakdown}. 
A composite mixture distribution for the 10nm node, shown in Fig \ref{fig:ci}, was created by performing KDE on both South Korea's and Taiwan's historical electric grid carbon intensity data, with weights proportional to the percentage of production capacity.

While the aforementioned uncertainty modeling focuses on the carbon intensity of semiconductor fabs, the same process can be applied to the carbon intensity during operational use, $\text{CI}_{use}$. 
Section~\ref{sec:operationalCI} evaluates both embodied and operational uncertainty for example data center scale and mobile scale processors.

\subsection{Evaluating End-to-end Uncertainty}
\begin{figure}
    \centering
    \includegraphics[width=\linewidth]{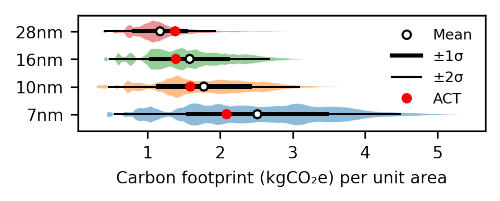}
    \vspace*{-8mm}
    \caption{Embodied carbon emission per unit area (cm$^2$) with its uncertainty distribution for technology nodes 7nm, 10nm, 16nm and 28nm under mass mature production. ACT's~\cite{act} deterministic point estimates are added for comparison.}
    \label{fig:all-nodes}
    \vspace{-1 mm}
\end{figure}

In this section we put the individual sources of uncertainty (i.e., EPA, yield, GPA, carbon intensity of the fab) together to analyse the overall impact on carbon-per-unit area estimates.
Fig.~\ref{fig:all-nodes} illustrates the embodied carbon emissions per cm$^2$ for four technology nodes, 7nm, 10nm, 16nm, and 28nm.
The average embodied carbon increases from 1.18 kgCO$_2$ per cm$^2$ to 2.52 kgCO$_2$ per cm$^2$ between older 28nm process node to newer 7nm process node.
\tool~allows designers to consider not only the average emissions but also the uncertainty.
Designers may select a percentile from the carbon footprint distribution based on system requirements and sustainability reporting standards.
For instance, compared to the average 1.18 kgCO$_2$ per cm$^2$ footprint for 28nm, the first standard deviation and 95$^{th}$ percentile emissions are 1.56 and 1.92 kgCO$_2$ per cm$^2$, respectively. 
We observe the impact of uncertainty is more pronounced on newer process technologies.
For instance, the standard deviation in carbon footprint estimates is 0.38, 0.56, 0.66, and 0.99 kgCO$_2$ for 28$nm$, 16$nm$, 10$nm$, and 7$nm$.
As technology nodes advance, accounting for both the mean and uncertainty in carbon estimates becomes increasingly important.
For instance, the 95$^{th}$ percentile carbon estimate for 7nm is more than 1.6$\times$ higher (4.13 kgCO$_2$) than the average estimate (2.52 kgCO$_2$).

In the following section, we describe how designers can use \tool\text{}~to address uncertainty in embodied carbon estimates.



\section{System Sustainability Assessment}

\begin{figure}
    \centering
    \includegraphics[width=\linewidth]{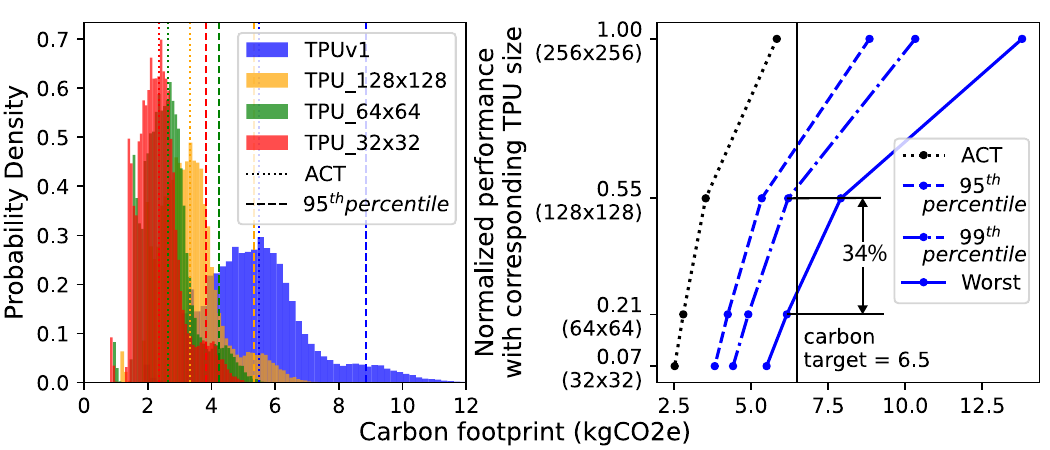}
    \vspace*{-6mm}
    \caption{The impact of uncertainty on provisioning resource for an AI accelerator. The left figure shows the embodied carbon footprint uncertainty distribution for TPU with various systolic array sizes. The right-side figure shows given a carbon footprint target, choosing TPU size of 128x128 can offer the performance increase by 34\% comparing to worst-case estimation and also keeping the risk of exceeding the carbon target to less 0.05\%.}
    \label{fig:TPU}
    \vspace{-2mm}
\end{figure}

\subsection{System Provisioning Using TPUs as An Example}
\textbf{Takeaway 1:} \textit{Uncertainty-aware provisioning enables hardware design choices that better align with both carbon reduction commitments and system performance goals. It avoids the excessive conservatism of worst-case deterministic estimates and the risks of overly optimistic average-case deterministic estimates.}

Every year companies like Apple and Google need to reduce their carbon emission by 6 to 10\% in order to hit their net-zero carbon emission goal by 2030~\cite{appleSustainabilityReport, googleEnvironmentalReport}. ACT~\cite{act} shows 74\% of these companies' carbon emissions come from manufacturing. 
As a result, companies face two contrasting approaches when estimating embodied carbon emissions: i) relying on the worst-case scenario, which can result in overly pessimistic estimates, leading to excessive conservatism~\cite{googleEnvironmentalReport}; ii) using the ACT value, which aligns with the mean of the probabilistic distribution.  This may lead to overly optimistic estimates, underestimating potential embodied carbon during IC design and acquisition.

To help companies navigate this trade-off, we show that \tool's probabilistic framework can enable more informed decision-making by balancing carbon estimation conservatism against performance targets. Rather than optimizing purely for the worst-case or average case, designers can strategically select designs that achieve high performance while maintaining a low risk of exceeding carbon targets. We illustrate this through a resource provisioning scenario based on AI accelerator TPUs \cite{tpuv1}.

Specifically, we explore how companies might choose between different TPU sizes (systolic array sizes of 32×32, 64×64, 128×128, and 256×256) depending on their performance need and carbon target. TPUv1, with a 256×256 systolic array and an area of 331mm², serves as our baseline. To estimate smaller designs, we scale the systolic array area quadratically and buffer area linearly based on side length. We simulate the performance of these designs using SCALE-Sim~\cite{SCALESIM} running DeepBench~\cite{deepbench}. The power estimation is based on synthesis results from 3D IC\cite{3D-IC}. 

Fig. \ref{fig:TPU} (left) shows that a smaller systolic array leads to a reduced area, which in turn lowers the uncertainty in the embodied carbon distribution. The variance decreases from 2.6 (kgCO$_2$)$^2$ for a 256x256 systolic array to 0.47(kgCO$_2$)$^2$ for a 32x32 array, while simultaneously reducing overall embodied carbon emissions.  Fig. 8 (right) evaluates performance, area, and carbon trade-offs based on the embodied carbon uncertainty distribution. While the 256×256 array delivers the highest performance (normalized to 1.0), the 128×128 array achieves a more balanced design: 0.55 normalized performance, 46.9\% reduction in power, 30\% reduction in area, and 43\% lower embodied carbon emissions under the 95$^{th}$ percentile carbon estimation.

Fig. 8 (right) also illustrates that while reducing the array size decreases embodied carbon and its uncertainty, it leads to diminishing performance returns. Companies optimizing purely for worst-case carbon compliance would prefer the 64×64 design, as it minimizes risk but comes at a major performance cost. In contrast, using the full uncertainty distribution enables companies to adopt a more balanced strategy: selecting a 128×128 TPU size improves performance by 34\% compared to the conservative 64×64 option, while keeping the probability of exceeding the carbon target below 5\%. Conversely, relying on average-case estimates alone and choosing the 256×256 design would risk a 21.2\% carbon overrun. Thus, uncertainty-aware modeling helps companies provision hardware aligned with both performance and sustainability goals..

\begin{figure}
    \centering
    \includegraphics[width=\linewidth]{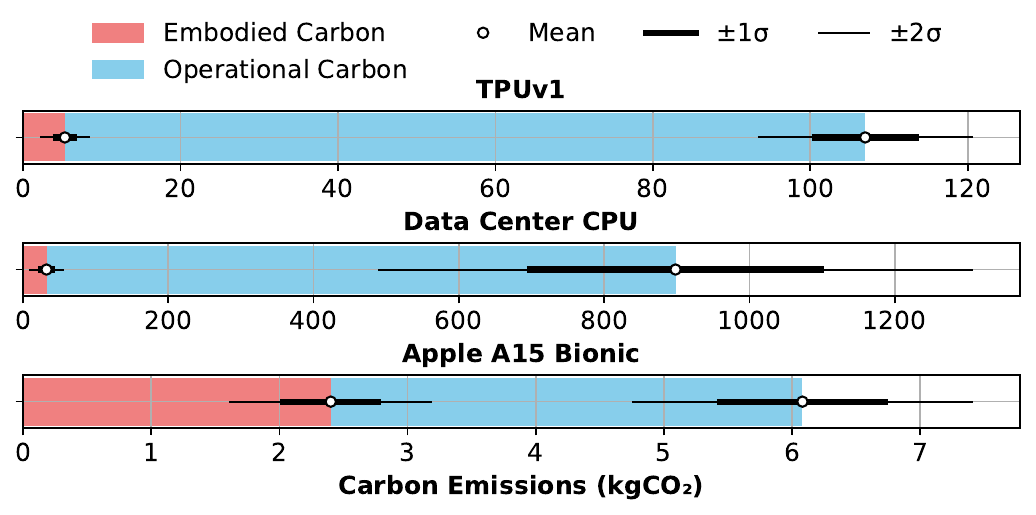}
    \vspace*{-6mm}
    \caption{Total carbon emissions of TPUv1, Data Center CPU AMD EPYC 7763v (Milan), and Apple A15 Bionic, broken down into embodied and operational carbon. Uncertainty in both embodied and operational carbon is indicated using standard deviations. Operational carbon emissions are estimated using the U.S. average carbon intensity. While TPUv1 and the data center CPU are operational carbon dominated under this energy mix, the Apple A15 exhibits a relatively balanced split between embodied and operational emissions.}
    \vspace{-1mm}
    \label{fig:total-carbon}
\end{figure}

\begin{figure}
    \centering
    \includegraphics[width=\linewidth]{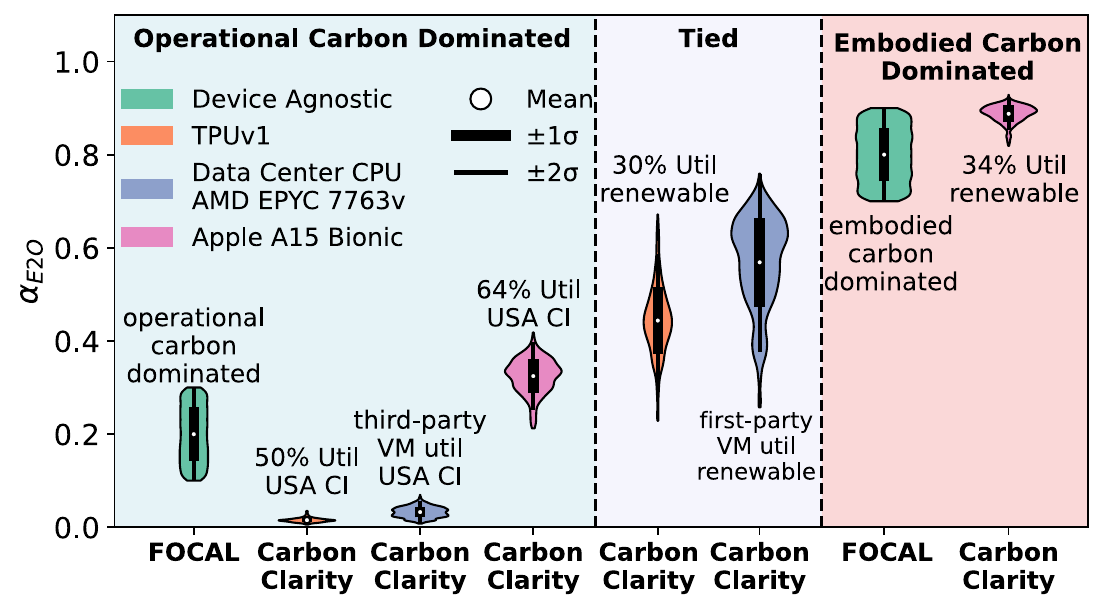}
    \vspace*{-6mm}
    \caption{Distribution of embodied-to-operation weight ($\alpha_{E2O}$) across different hardware, utilization scenarios, and energy mixes.
The plot compares $\alpha_{E2O}$ distributions generated by FOCAL and CarbonClarity. FOCAL assumes a binary classification into embodied carbon dominated ($\alpha = 0.8 \pm 0.1$) and operational carbon dominated ($\alpha = 0.2 \pm 0.1$) regions, resulting in broad uniform distributions. In contrast, CarbonClarity models $\alpha_{E2O}$ probabilistically, capturing the impact of real-world variability in utilization and carbon intensity. Results show that the same hardware can span a wide range of $\alpha_{E2O}$ values depending on deployment conditions, highlighting the importance of scenario-specific sustainability assessments based on quantitative uncertainty modeling.}
    \vspace{-1mm}
    \label{fig:alpha}
\end{figure}

\subsection{Adding Operational Carbon Uncertainty to the Equation: Comparison with FOCAL}~\label{sec:operationalCI}

\textbf{Takeaway 2:} \textit{We highlight the need to quantitatively account for uncertainty in sustainability assessments, both in embodied and operational carbon. Inherent variability in workload, energy mix, and utilization can shift the carbon footprint of a system from embodied to operational, a behavior that binary assumptions cannot capture.}

Although the primary focus of \tool~is to quantify the uncertainty of embodied carbon emissions for SoCs, the same principles of uncertainty modeling can and should be extended to operational carbon emissions by considering variability in usage patterns and energy mix.
Figure~\ref{fig:total-carbon} shows the estimated total carbon emissions, both operational and embodied, for TPUv1, a data center CPU (AMD EPYC 7763v, Milan) and a mobile Apple A15 Bionic Chip. 
To model operational carbon uncertainty, we identify two major empirical factors: device utilization and energy source carbon intensity (CI).
For TPUv1, operational carbon emissions are estimated by using reported power consumptions\cite{tpuv1} with a projected deployment lifetime of five years, applying the utilization distribution of AI accelerators as reported by production GPU utilization in Meta's data centers~\cite{sustainableAIpaper}. 
For the data center CPU, operational carbon emissions are approximated using the device’s thermal design power (TDP) and projected over a five-year deployment based on CPU utilization data from Microsoft's cloud data centers~\cite{ResourceCentral}. 
For the mobile bionic SoC, the operational emissions are estimated using the chip TDP~\cite{A15tdp} and utilization~\cite{mobileUtil} based on a deployment time of 2.59 years\cite{mobileLifetime}.
In all three cases, operational energy consumption is converted into carbon emissions using the US average carbon intensity for the data presented in Fig.~\ref{fig:total-carbon}.

As shown in Fig.~\ref{fig:total-carbon}, operational carbon emissions dominate for the TPUv1 and CPU, with a 95th percentile operational footprint nearly 13.64$\times$ and 21.95$\times$ higher than its 95th percentile embodied footprint.
On the other hand, the mobile CPU shows a more balanced profile where 95th percentile embodied and operational carbon are roughly equal.

Given the availability of uncertainty modeling for both embodied and operational carbon emissions, their relative importance becomes critical for system-level sustainability assessments.
Under the current state-of-the-art, this tradeoff has been modeled using the \textit{embodied-to-operation (E2O) weight} $\alpha_{E2O}$ introduced by FOCAL. 
FOCAL assigns a fixed weight depending on whether a system is assumed to be dominated by embodied carbon ($\alpha = 0.8 \pm 0.1$) or dominated by operational carbon ($\alpha = 0.2 \pm 0.1$), based on previous work~\cite{chasingcarbon}.
FOCAL’s analytical model results in a uniform distribution over its assumed range, failing to capture the true, highly scenario-dependent variability and real-world behavior.

As shown in Fig.~\ref{fig:alpha}, even for the same hardware, $\alpha_{E2O}$ values vary substantially based on utilization and energy mix. 
For example, under higher utilization and higher carbon intensity energy sources, all three processors (that is, TPUv1, Apple A15 Bionic SoC, and AMD CPU) fall within the operationally dominated domain. 
However, while TPUv1 and AMD CPU are below FOCAL's lowest assumption of $\alpha_{E2O}$ of 0.1, the Apple A15 Bionic has a 95th percentile $\alpha_{E2O}$ of 0.38, showcasing a near parity between operational and embodied carbon.
Furthermore, both TPUv1 at 30\% utilization and AMD CPUs under first-party utilization (approximately 15.97\%~\cite{ResourceCentral}) with renewable energy sources yield $\alpha_{E2O}$ values indicating a tie between operational carbon dominance and embodied carbon dominance.
Designers using FOCAL's preset regions of operational and embodied dominated regimes would not consider these scenarios, when in reality designers looking to deploy sustainable hardware would need to carefully consider operational and embodied overheads of these systems.

Finally, considering different typical utilization and carbon intensities, $\alpha_{E2O}$ for Apple A15 Bionic SoC changes from operational dominated to embodied dominated regimes.
This highlights the need for sustainability modeling frameworks that goes beyond binary classifications based on a priori knowledge of the life cycle emissions of systems (i.e., $\alpha_{E2O}$), offering a richer probabilistic understanding of the carbon profile of a system under various real-world conditions.
\section{Chiplet Design Case Studies}
We now show the implications of the probabilistic model have on system provisioning and chiplet designs, compared to a baseline deterministic model, ACT~\cite{act}.
We identify key contributors to carbon estimate uncertainty and propose strategies to mitigate embodied carbon variability.

\begin{figure}
    \centering    \includegraphics[width=\linewidth]{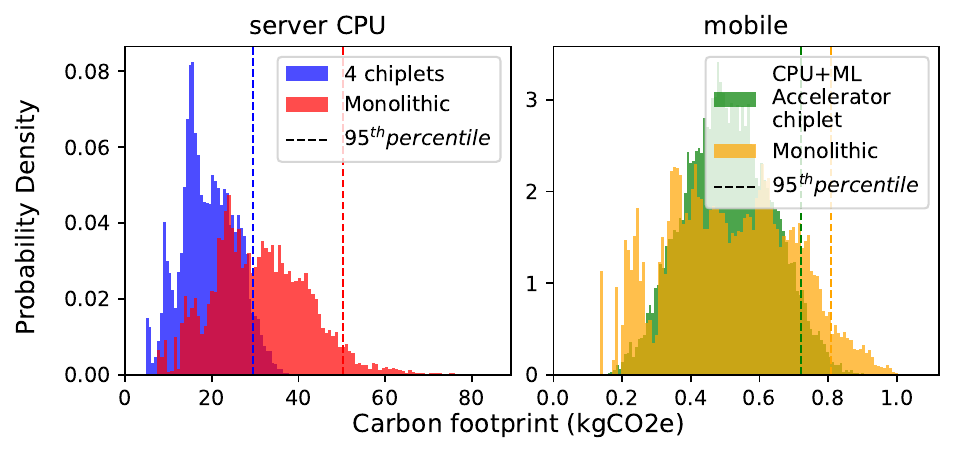}
    \vspace*{-7mm}
    \caption{chiplet vs monolithic uncertainty comparison for both AMD EPYC$^\text{TM}$ and mobile with ML accelerator incoporated. The 95$^{th}$ percentile estimation, represented by dashed lines, indicates the recommended embodied carbon value for carbon estimation.}
    \label{fig:chiplet}
    \vspace{-1mm}
\end{figure}

\begin{figure}
    \centering
    \includegraphics[width=\linewidth]{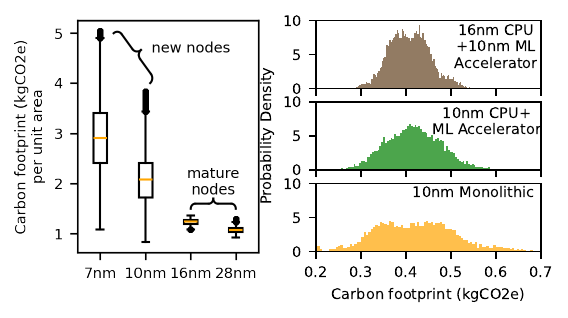}
    \vspace*{-8mm}
    \caption{The left figure shows the uncertainty distribution for mature technology nodes (16nm and 28nm) compared to newer technology nodes (7nm and 10nm) that entered mass production between 2017 and 2019. The right figure illustrates a mix of new and old technology nodes in a small mobile chiplet design. The reduced embodied carbon and lower uncertainty of the older technology node (16nm) help offset the impact of area scaling, making it a more sustainable choice than both monolithic designs and chiplet designs using a newer technology node.}
    \label{fig:new_vs_old}
    \vspace{-1mm}
\end{figure}

\begin{figure*}
    \centering
    \includegraphics[width=\linewidth]{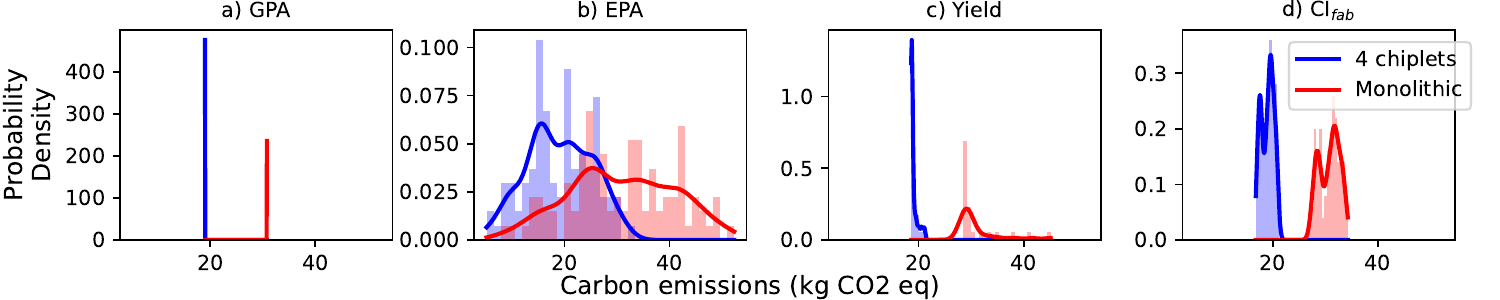}
    \vspace*{-6mm}
    \caption{distribution for each source of uncertainty. This is an uncertainty source analysis of the AMD EPYC$^\text{TM}$~\cite{AMD_EPYC}. EPA has the highest uncertainty out of all uncertainty sources. Yield uncertainty level is dependent on the area of the chip.}
    \label{fig:uncertainty-source}
    \vspace{-8mm}
\end{figure*}

\subsection{Case Study I: Reducing Uncertainty with Chiplets}

\textbf{Takeaway 3:} \textit{Chiplet architectures not only reduce mean embodied carbon but also significantly tighten embodied carbon uncertainty distributions, mitigating the long-tail yield uncertainty associated with large monolithic chip designs.}

With the ability to reduce embodied carbon~\cite{AMDreport, ecochip}, chiplet architectures have become an increasingly popular option for IC design. In this section, we demonstrate that chiplets not only lower embodied carbon but also reduce embodied carbon uncertainties. We look at two scenarios: AMD EPYC$^\text{TM}$ server CPU~\cite{AMD_EPYC} and mobile CPU with ML accelerator incorporated. 

Fig. \ref{fig:chiplet} shows the embodied carbon distribution of the scenarios mentioned above. To construct the chiplet-based AMD EPYC$^\text{TM}$ design, we take the reported data of the monolithic chip size being 777mm$^2$ of a 32-core CPU, split into 4 chiplets size of 213mm$^2$, each with 8 cores, manufactured in 14nm process technology~\cite{AMD_EPYC}. The hypothetical mobile example assumes the monolithic design with an area of 20 mm$^2$, split to 10 mm$^2$ CPU chiplet and 10 mm$^2$ ML accelerator, manufactured in 10nm. 

We observe that splitting up a single monolithic chip into heterogeneous chiplets helps reduce the overall embodied carbon uncertainty, regardless of the applications. For example, in the mobile scenario shown in Fig. \ref{fig:chiplet} (right), while the mean embodied carbon remains similar, the 95$^{th}$ percentile carbon estimate is significantly reduced from 0.8 kgCO$_2$ to 0.72 kgCO$_2$, indicating a tighter carbon distribution. A larger chip area generally enlarges the uncertainty presence in GPA, EPA and yield. As shown in Fig. \ref{fig:chiplet} (left), the large chip area in server-level chips can have a notable secondary-level effect on yield, as indicated by the long-tailing effect in AMD EPYC$^\text{TM}$'s monolithic distribution (shown in red). The chiplet configuration demonstrates a significant reduction in embodied carbon uncertainty, with a 95$^{th}$ percentile embodied carbon estimate of 29.4 kgCO$_2$ compared to 50.67 kgCO$_2$ for the monolithic design.

\vspace{-1mm}
\subsection{Case Study II: Chiplet New vs Old Technology Nodes}

\textbf{Takeaway 4:} \textit{Combining mature and advanced technology nodes in chiplet-based designs reduces both embodied carbon and its uncertainty, exploiting the lowered carbon distribution variability and improved yield of mature nodes to counterbalance the higher uncertainty of newer technologies.}

As each process node advances into mass production, its energy-per-area (EPA) and yield improve with time~\cite{TSMCepa, defectDensity}. To illustrate these trends, we plot the embodied carbon distribution across various technology nodes (7nm, 10nm, 16nm, and 28nm), highlighting both the changes in embodied carbon and the associated uncertainty levels. Fig. \ref{fig:new_vs_old}
(left) shows the carbon emissions per cm² for the 7nm, 10nm, 16nm, and 28nm nodes from 2017 to 2019. During this period, the 16nm and 28nm nodes were mature, while the 7nm and 10nm nodes had just entered mass production in 2017.

Our analysis reveals that newer nodes (7nm and 10nm) exhibit significantly higher uncertainty than the more mature nodes (16nm and 28nm). This is largely due to the initial inefficiencies and rapid improvements in EPA shortly after mass production begins~\cite{TSMCepa}. As EPA gradually stabilizes and reaches a plateau, the uncertainty decreases. For example, the variances for the newer nodes are 0.51 (kgCO$_2$)$^2$ and 0.32 (kgCO$_2$)$^2$ for 7nm and 10nm, respectively, reflecting greater uncertainty. However, the variances of the more mature 16nm and 28nm nodes are 0.19 (kgCO$_2$)$^2$ and 0.13 (kgCO$_2$)$^2$. Despite these improvements in uncertainty, the average embodied carbon per cm$^2$ tends to increase with newer technology nodes.

We apply the mix-and-match approach of new and old technology nodes to our mobile chiplet study, as shown in Fig. \ref{fig:new_vs_old} (right). The 10nm monolithic chip (20mm$^2$), which includes both the CPU and ML accelerator, is represented by the yellow distribution. The green distribution represents a configuration with a 10nm 10mm$^2$ CPU chiplet and a 10nm 10mm$^2$ ML accelerator. The brown distribution shows a 16nm CPU chip with area scaling~\cite{areaScaling} and a 10nm ML accelerator. The combination of the mature 16nm node with the newer 10nm node results in the lowest embodied carbon and carbon uncertainty, with its 95$^{th}$ percentile value 7.5\% lower than the 10nm chiplet configuration (green) and 18\% lower than the monolithic design (yellow). This demonstrates that the mix-and-match approach of new and old technology designs not only effectively reduce embodied carbon but also lower the uncertainty in carbon estimates.

\subsection{Case Study III: Identifying Key Sources of Uncertainty}

\textbf{Takeaway 5:} \textit{Yield variability and energy-per-area (EPA) fluctuations are the dominant contributors to uncertainty. The high EPA variability underscores the need for greater adoption of renewables in semiconductor fabrication. Fabs can leverage \tool~to pinpoint key sources of uncertainty and strategically prioritize uncertainty reduction efforts.}

In addition to reducing uncertainty, analyze chiplet designs using~\tool~can help designers identify sources of uncertainty, enabling future improvements and mitigation strategies. To determine the extent to how each source of uncertainty influences the overall embodied carbon distribution, we plot the uncertainty distribution of EPA, GPA, Yield and CI$_{fab}$ from AMD EPYC$^{\text{TM}}$ chiplet vs monolithic in Fig. \ref{fig:uncertainty-source}. To examine each individual uncertainty source, we held all other sources constant at the mean of their distributions. Our findings revealed several key points.

Yield uncertainty creates a long-tail effect for large-area chips, as shown in Fig. \ref{fig:uncertainty-source}c. Larger chips not only generate more embodied carbon but also exhibit greater uncertainty in their emissions. The long-tail effect of yield helps explain why the chiplet design in Case Study I effectively reduces its carbon emissions, especially compared to large monolithic chips.

EPA values introduce the highest uncertainty in the embodied carbon uncertainty model, indicated by Fig.~\ref{fig:uncertainty-source}b. The large variance in EPA, when amplified by the high carbon intensity of non-renewable energy sources, can significantly increase the uncertainty in the overall embodied carbon distribution. As a result, powering the fab with more renewable energy sources not only reduces embodied carbon but also decreases the uncertainty in the embodied carbon estimates.


GPA, on the other hand, contributes the least to the overall embodied carbon uncertainty. This is partly because 87\% of embodied carbon emissions are dominated by CO$_2$ from energy use~\cite{TSMC2023report}. Therefore, while GPA uncertainty is non-trivial, its impact on the total embodied carbon uncertainty is relatively limited at this stage, especially when non-renewable energy sources still dominate the energy used in fabs.

\vspace{-1mm}
\section{Conclusion and Future Works}
We presented \tool, a probabilistic framework to address uncertainty in embodied carbon estimates, enabling risk-aware and sustainable architecture design decisions. We show that despite the high uncertainty in carbon models, designers can use our framework to understand and design around uncertainty. Through a series of case studies, we demonstrate the importance of incorporating uncertainty into resource provisioning and highlight how chiplet-based architectures can effectively mitigate uncertainty. Future efforts can extend this framework by incorporating additional supply chain dynamics and exploring how emerging technologies can be modeled to support sustainability-aware decision making.
\vspace{-1mm}
\section{Acknowledgment}
This research was supported by the National Science Foundation under award numbers CCF-2324860 and CNS-2335795.

\clearpage
\begingroup
\bibliographystyle{IEEEtran}
\bibliography{ref}

\end{document}